# Affective visualization in Virtual Reality: An integrative review

Andres Pinilla[1, 2 *], Jaime Garcia[2], William Raffe[2], Jan-Niklas Voigt-Antons[1, 3], Robert P. Spang[1], Sebastian Möller[1, 3]

[1] Quality and Usability Lab, Institute for Software Technology and Theoretical Computer Science, Faculty of Electrical Engineering and Computer Science, Technische Universität Berlin, Germany

[2] UTS Games Studio, Faculty of Engineering and IT, University of Technology Sydney UTS, Sydney, Australia

[3] German Research Center for Artificial Intelligence (DFKI), Berlin, Germany

*** Correspondence:**
Andres Pinilla
andres.pinilla@qu.tu-berlin.de

## Abstract

A cluster of research in Affective Computing suggests that it is possible to infer some characteristics of users' affective states by analyzing their electrophysiological activity in real-time. However, it is not clear how to use the information extracted from electrophysiological signals to create visual representations of the affective states of Virtual Reality (VR) users. Visualization of users' affective states in VR can lead to biofeedback therapies for mental health care. Understanding how to visualize affective states in VR requires an interdisciplinary approach that integrates psychology, electrophysiology, and audio-visual design. Therefore, this review aims to integrate previous studies from these fields to understand how to develop virtual environments that can automatically create visual representations of users' affective states. The manuscript addresses this challenge in four sections: First, theories related to emotion and affect are summarized. Second, evidence suggesting that visual and sound cues tend to be associated with affective states are discussed. Third, some of the available methods for assessing affect are described. The fourth and final section contains five practical considerations for the development of virtual reality environments for affect visualization.

Virtual Reality (VR) systems offer endless possibilities for the development of interactive experiences. They are used for the development of tools in diverse areas such as rehabilitation therapy (Garcia & Navarro, 2014), exergames (Arndt et al., 2018; Greinacher et al., 2020), and robotics (Burdea et al., 2013). Their potential is particularly promising when combined with technological advances in Affective Computing, allowing to interpret users' affective states as computer commands (Hernandez et al., 2014; Leslie et al., 2015; Picard et al., 2001; Sitaram et al., 2011) and adapt the content of a virtual environment accordingly (Bermudez i Badia et al., 2019).

Traditional psychological tasks for the treatment and diagnosis of mental disorders can be replaced by VR systems (Belger et al., 2019; J. Blum et al., 2019, 2020; Koenig et al., 2011). These new tools are less time-consuming and provide more realistic environments, hence higher ecological validity. Furthermore, VR might be helpful for at least two types of therapy: exposure therapy and biofeedback therapy. Exposure therapy is commonly used to treat anxiety disorders caused by phobias. Patients are systematically exposed to the stimuli that trigger the phobia in a controlled environment and with a therapist's guidance. VR is useful for exposure therapy because it allows delivering realistic experiences while providing control over the stimuli. Previous research suggests that exposure therapy in VR might be effective for the treatment of a least three types of phobias: social phobia (Shiban et al., 2015), claustrophobia (Shiban et al., 2016), and spider-phobia (Peperkorn et al., 2014).

Biofeedback therapy is used to provide real-time feedback to the patient about their physiological activity while they perform a task (J. Blum et al., 2020). The characteristics of the task depend on the purpose of the therapy. For example, Blandón et al. (2016) developed a biofeedback game for training attention control in children with Attention Deficit Hyperactivity Disorder (ADHD). The player performed tasks in a virtual farm, such as collecting fruits and repairing a pathway. Participants were challenged to increase their concentration, impulsivity control, and sustained attention to do these tasks. Simultaneously, electroencephalography (EEG) signals were processed in real-time to identify EEG activity associated with attention state. The player obtained a game score if attention state was detected in the EEG signals.

Similarly, Cavazza et al. (2014) developed an interactive experience to enhance empathy using neurofeedback. The participant interacted with a fictional doctor that was going through a difficult situation. Simultaneously, the participant's EEG signals were analyzed to estimate the affective response towards the doctor. If the system detected a positive affective response in the player, the storyline would change positively (the doctor would struggle less). It was expected that those changes in the story line would reinforce the supportive, empathetic behavior of the player.

Patients who lack affective self-regulation could benefit from VR biofeedback therapy to train affective self-control, fostering mood regulation (Desmet, 2015). Li et al. (2016) conducted an experiment where twenty-three participants' brain activity was analyzed in real-time using functional Magnetic Brain Imaging (fMRI). They asked participants to evoke a happy or sad memory and provided feedback about their affective state. The feedback was provided with a bar on a screen. The bar's level increased when the fMRI data indicated that the participant successfully evoked the happy or sad memory. Results suggest that providing visual feedback allowed participants to learn how to modulate their neural activity. But it is not clear how to implement this finding in a therapeutic application that can be accessible for a large population because (1) fMRI is an expensive technology that is not accessible for most people, (2) participants must remain motionless for long periods during fMRI recordings; otherwise, the data is corrupted, and (3) an ideal therapeutic application should consist of engaging content that motivates users to use the system. These challenges could be solved by measuring brain activity with a less expensive and more portable method than fMRI, such as electroencephalography (EEG). The visual feedback could be provided using game-like elements.

The development of an affective visualization tool in VR would require at least two components: (1) A set of VR stimuli with affective content whose properties can be adjusted in real-time, and (2) a technique to continuously assess affective states in an online fashion without interrupting the VR experience. This literature review was elaborated to understand how to develop those components. Both requirements are addressed in three subsections. Firstly, theories related to emotions and affect are presented. Secondly, findings related to visual and sound cues that are associated with affective responses are analyzed. And thirdly, some of the most common methods for detecting affective states are summarized.

## 1 Theoretical models of emotion and affect

The terms emotion and affect are often used interchangeably in the literature, but they are not exactly the same. There is not a general agreement about how to define these concepts. In this manuscript, emotions are defined as mental states that coordinate the operation of cognitive processes. This definition is based on the assumption that the human mind is designed as a computational system that consists of a series of information-processing programs (Putnam, 1967). Emotions are a particular type of program that coordinate other programs' operation (Cosmides & Tooby, 1994). Affect is defined as the cognitive representation of the bodily changes that come with emotions (Barrett & Bliss-Moreau, 2009; Wundt, 1897). Neither emotion nor affect can be directly observed or measured. However, affect is conceptually associated with the physiological changes of the body. Therefore, it is reasonable to use electrophysiological signals to infer affective states, which might allow to infer some characteristics of emotional states.

### 1.1 Emotion theories

The Ortony, Clore & Collins (OCC) theory of emotions (Ortony et al., 1988) has been widely used in the field of computer science to model users' emotional responses (e.g., Conati & Zhou, 2002; Jaques & Vicari, 2007). This theory describes emotions in terms of twenty-two categories and assumes a clear distinction between each category. This approach is compatible with existing emotion recognition algorithms because these are usually based on categorizing emotions (e.g., Harischandra & Perera, 2012; Mavridou et al., 2017). According to the OCC theory (Ortony et al., 1988), the first step in an emotional response is the perception of the situation. Then the situation is evaluated (appraisal), and finally, the emotional response emerges. However, this theory does not consider the physiological changes associated with emotions.

Similarly, Robert Plutchnik proposed a structural model of emotions (Plutchik, 1982), commonly known as Plutchnik's wheel of emotions. This model consists of eight primary states: ecstasy, adoration, terror, amazement, grief, loathing, rage, and vigilance. According to Pluthnik's theory, any emotion can be described as a combination of a subset of those basic states. Here emotions are defined as a sequence of reactions towards a stimulus. This sequence includes a cognitive evaluation of the stimuli (appraisal), feelings (subjective experience of the emotion), autonomic neural activity, and behavioral responses.

There are at least three other major emotion theories: the James-Lang Theory (Lange & James, 1922), the Cannon-Bard Theory (Bard, 1934; Cannon, 1927), and the Schachter-Singer Theory (Schachter & Singer, 1962). According to Shiota & Kalat (2012), these theories have in common the assumption that emotional responses have four components but differ in the order those components take place during an emotional response. The components are:

- **Appraisal:** The cognitive, rationalized evaluation of the context where the emotional response is produced.
- **Feeling:** The subjective, momentary experience of the emotion.
- **Physiological change:** The bodily changes produced by the emotional response.
- **Behavior:** The observable conduct that comes with the emotion.

According to the **James-Lang Theory** (Lange & James, 1922), the first step in an emotional response is the cognitive evaluation of the situation. Then, physiological changes are produced in the body, at the same time that a behavioral response is produced. Lastly, feelings take place.

The **Cannon-Bard theory** (Bard, 1934; Cannon, 1927) proposes that all the elements of an emotional response are independent of each other, and there is no particular order in which they occur. This theory is not compatible with the convincing amount of evidence indicating that emotional stimuli tend to trigger automatic changes in the body (e.g., Dimberg et al., 2000; Huster et al., 2009; Thayer et al., 2009). Overall, these previous studies suggest interdependence between physiological changes and the other components of emotion.

According to the **Schachter-Singer theory** (Schachter & Singer, 1962), physiological changes occur first. Then the user tries to find an explanation in the environment for those physiological changes. Depending on the explanation found, a label is assigned to the bodily changes perceived. Therefore, the physiological changes indicate the intensity of the emotional experience, but cognitive factors determine the emotion's valence (pleasant vs. unpleasant).

## 1.2 Theoretical models of affect

Theoretical models of affect can be classified into two major groups: discrete and dimensional models. Discrete models are based on a categorical division of affective responses, while dimensional models represent affect as an array of continuous variables. Both types of models are commonly used in Affective Computing to build affect recognition models (e.g., Hernandez et al., 2014; Leslie et al., 2015; Sitaram et al., 2011).

In broad terms, discrete models propose the existence of a few primary states, such as happiness, sadness, and anger. Affective responses are a combination of a subset of those fundamental states. Evidence obtained by Ekman and Friesen (1971) during an experiment conducted in New Guinea supports discrete models. In this experiment, stories with emotional content were told to 153 participants. One hundred thirty of them (84.97%) had no previous contact with the western culture. After each story had been told, participants saw a series of pictures of facial expressions and were asked to choose the more coherent face with the story. Interestingly, participants associated similar facial expressions to the same stories, regardless of their cultural background. Based on this evidence, it was proposed that there are at least six facial expressions that are universal (i.e., they are not affected by culture): happiness, anger, sadness, disgust, surprise, and fear. These results are consistent with earlier contributions from Charles Darwin, who pointed out the existence of activation patterns in facial muscles which are associated with affective states (Darwin, 1872; Ekman, 2006).

Dimensional models have their roots in the early contributions of Wilhelm Wundt, who proposed that affective responses have three dimensions: valence (pleasant – unpleasant), arousal (arousing – subduing), and intensity (strain – relaxation) (Wundt, 1897). On this basis, the Circumplex Model of Affect (Russell, 1980) was developed, representing affect in a two-dimensional space, where valence and arousal are equivalent to the *x*-axis and *y*-axis, respectively.

Other authors have proposed the Evaluative Space Model (ESM) (Cacioppo et al., 1997), which has three dimensions: Negativity in the *x*-axis, Positivity in the *y*-axis, and Net predisposition (to withdraw or approach a stimulus) in the *z*-axis. Unlike the Circumplex Model of Affect (Russell, 1980), the ESM (Cacioppo et al., 1997) contemplates the existence of affective responses with simultaneous negative and positive activation (“bitter-sweet” affective states). For example, while playing a terror video game, the user might feel fear, and at the same time, might feel excited because there is not a real danger. An analysis about dimensional models of affect can be found in Mattek et al. (2017).

The ESM proposes the existence of the negativity bias and the positivity offset. The negativity bias implies that negative activation produces more changes in the motivation to withdraw or approach stimuli than positive activation. Evidence supporting the existence of the negativity bias indicates that negative stimuli tend to produce more salient behaviors than positive stimuli (Sutherland & Mather, 2012), and negative stimuli tend to be associated with higher arousal than positive stimuli (Lang et al., 2008). The negativity bias suggests that terror video games should trigger higher arousal than video games associated with positive affective states. However, a recent study indicates that the arousal level triggered by terror video games is slightly lower than the arousal triggered by video games associated with positive affective states (Martínez-Tejada et al., 2021).

The positivity offset implies a slight positive motivation to approach unknown stimuli in a neutral environment. This mechanism has been associated with humans' natural tendency to explore new, unthreatening environments, even when that behavior is not associated with a reward (Cacioppo et al., 1997, p. 12). Further research about the positivity offset could help understand how to motivate VR users to explore virtual environments. For example, to stimulate engagement of players with VR games.

## 2 Visual and sound cues

Building a virtual environment for affective visualization requires content that any user can associate with a wide range of affective states, regardless of cultural differences or personal preferences. Therefore, this section presents recent studies suggesting an association between some characteristics of graphical elements and affective states. We do not intend to define a set of rules about how to communicate affect with audio-visual elements. Instead, we aim to provide guidelines about visual and sound features when creating visual representations of affective states.

### 2.1 Visual cues

Rounded objects are associated with higher valence and lower arousal than sharp objects (Bar & Neta, 2006). And rounded lines are perceived as more attractive than straight or angular lines (Aronoff, 2006; Aronoff et al., 1992). Given that attractiveness is associated with positive affective states, rounded lines are likely associated with positive valence. Additional studies suggest that visual complexity plays a role in the likability of objects. People tend to prefer extremely simple or extremely complex objects (Norman et al., 2010). Given that likability tends to trigger positive valence (Ryali et al., 2020), an intermediate level of complexity is more likely associated with negative valence.

A cross-cultural study showed that the most critical factors in the affective meaning of color are brightness and saturation, while hue has a secondary role (Gao et al., 2007). These results are consistent with evidence reported in Valdez & Mehrabian (1994) but contrast with recent studies indicating that hue has a significant role in the affective state associated with a color palette (Bartram et al., 2017). Additional evidence suggests that blue, green, and purple are among the most pleasant hues, while yellow is among the most unpleasant. Green-yellow, blue-green, and green are the most arousing, while purple-blue and yellow-red are among the least arousing (Palmer & Schloss, 2010). Similarly, it has been found that the most pleasant colors are those with higher saturation and brightness (Camgöz et al., 2002; Wilms & Oberfeld, 2018). However, other studies suggest that there are not universal associations between colors and affective states. People tend to like colors associated with objects they like and dislike colors associated with objects they dislike (Palmer & Schloss, 2010). Additional evidence indicates that color associations change according to the context where colors are used (Lipson-Smith et al., 2020), supporting the hypothesis that there are not universal associations between colors and affective states. Yet, it is possible to establish color palettes that allow to communicate affective states. For example, bright, unsaturated colors are more suitable to communicate calm, while dark, red colors are more suitable to communicate disturbance (Bartram et al., 2017).

Textures may influence the affective meaning of color (Ebe & Umemuro, 2015; Lucassen et al., 2011). This has been demonstrated by pairing colors with computer-generated textures and asking participants to rate the color-texture pairs using four scales: Warm-Cool, Masculine-Feminine, Hard-Soft, and Heavy-Light. Results suggest that texture significantly influences the evaluation on the Hard-Soft scale and has a minor impact on the other scales. However, this evidence does not allow to identify associations between particular texture patterns and affective responses.

Non-static visual elements have other visual properties besides color, shape, and texture. Some of these additional properties are speed, motion shape, direction, and path curvature. Fast-moving objects are associated with higher arousal than slow-moving objects (Feng et al., 2014; Piwek et al., 2015). But there are contradictory findings regarding the type of valence associated with speed. One study suggest that fast movements are related to positive affective states (Piwek et al., 2015), while other study indicates the opposite (Feng et al., 2014).

Linear motion with straight paths is associated with low arousal and positive valence (Feng et al., 2014). Jerky paths are associated with higher arousal than straight paths in linear motion (Feng et al., 2014; Lockyer et al., 2011). But the curvature of the path has no incidence in affective associations when applied to spherical or radial motion (Feng et al., 2014). Inward movements are related to more positive affective states than outward movements (Feng et al., 2014). Downwards-right motion tends to be linked to positive states, while upwards-left motion tends to be associated with negative states (Lockyer et al., 2011). In general, angular paths are related to more negative affective states than linear paths (Lockyer et al., 2011). And spherical motion patterns tend to be associated with higher arousal than linear motion patterns.

## 2.2 Sound cues

Previous research indicates that the location of a sound source influences the affective states associated with that sound. When the user cannot see where the object is (outside of the field of view), it is often associated with more arousing affective states than when the user can see it (inside the field of view) (Drossos et al., 2015; Tajadura-Jiménez, Larsson, et al., 2010). Similarly, sounds located further away in the space are related to less arousing responses (Tajadura-Jiménez et al., 2008). The perception of an approaching sound is associated with more arousing responses than the perception of it moving away (Tajadura-Jiménez, Väljamäe, et al., 2010). These phenomena are likely to be linked to mechanisms enforced by evolution (Cosmides & Tooby, 1994). Our primitive ancestors had more chances to survive if they were aware of the most potentially dangerous objects, such as those they could not see, were closer to them, or were approaching them.

The reverberation of the sound, which is associated with space's size, can influence affective associations (Tajadura-Jiménez, Larsson, et al., 2010). Lower reverberation (smaller rooms) is linked to more pleasant states than higher reverberation (larger rooms). Perhaps, because the primitive human being was better protected from predators in closed spaces, leading to an evolutionary process that favors the activation of attentional resources when we are in open areas.

Other studies indicate that asking people to rate pictures with affective content while listening to the sound of a heartbeat can influence their affective evaluations, as well as their heart rate (Tajadura-Jiménez et al., 2008). Here, the sound of a heart rate faster than the listener's one tends to increase their heart rate, while a slower sound seems to relax the listener's heart rate. Therefore, playing a fast heartbeat in the background might be an effective way of representing an increase in arousal.

On the other hand, music is pivotal for affective visualization because it can contribute to create more immersive experiences. However, it is a vast topic that will not be fully covered in this manuscript. Yet, it is important to mention that tempo influences music's affective perception (Fernández-Sotos et al., 2016). Faster tempo tends to be associated with higher arousal ratings, while slower tempo tends to be associated with lower arousal ratings. To the extent of our knowledge, there is no evidence suggesting that tempo influences valence ratings.

Major and minor chords are associated with positive and negative affective states, respectively (Gerardi & Gerken, 1995). Similarly, dissonant harmonies tend to be strongly associated with anger, and to a lesser extent, with fear (Petri, 2009). And it is possible to compose music based on people's affective states (Williams et al., 2017). However, it remains an open question whether it is feasible to do it in real-time, based on the user's electrophysiological signals.

**Table 1.** Summary of audio-visual cues associated with affective states, according to previous studies.

| | | **High Arousal** | **Low Arousal** | **High (Positive) Valence** | **Low (Negative) Valence** |
|---|---|---|---|---|---|
| **Static visual cues** | Shape | Angular | Rounded | Rounded | Angular |
| | Lines | | | Rounded | Straight |
| | Hue | Green-yellow, blue-green, and green | Purple-blue and yellow-red | Blue, green, and purple | Yellow |
| | Saturation | N/A | N/A | Saturated | Unsaturated |
| | Brightness | N/A | N/A | Bright | Dark |
| | Visual complexity | N/A | N/A | Extremely complex or extremely simple | Neither complex nor simple |
| **Non-static visual cues** | Speed | Fast | Slow | Some studies suggest that fast motion is associated with positive valence, while others suggest the opposite. | |
| | Motion shape | Spherical | Linear | N/A | N/A |
| | Direction | N/A | N/A | Downwards-right, inward | Upwards-left, outward |
| | Path curvature | Jerky | Straight | Jerky | Straight |
| **Sound cues** | Source location | Outside field view | Inside field view | N/A | N/A |
| | Distance to the sound source | Near | Far | N/A | N/A |
| | Sound source movement | Approaching | Receding | N/A | N/A |
| | Heartbeat sound | Fast (above 100 bpm) | Slow (below 60 bpm) | N/A | N/A |
| **Music** | Tempo | Fast | Slow | N/A | N/A |
| | Harmony | N/A | N/A | Major scale | Minor scale |

### 2.3 Personalized affective visualizations

There might be individual differences in the affective states that each user associates with the same audio-visual stimuli. These individual differences could be amplified as a consequence of personal experiences. An ideal system for affective visualization should account for those individual differences, delivering personalized visual representations of affective states, similar to Bermudez i Badia (2019).

Semertzidis et al. (2020) developed an Augmented Reality (AR) system that automatically creates visual representations of the user's affective states. The visualizations consisted of fractals generated using Procedural Content Generation (PCG). The visual properties of the fractals varied according to the affective state detected in the user. However, the evidence reported by Semertzidis et al. (2020) does not allow to establish whether participants perceived that the fractals' graphical properties represented their affective states.

Additional studies indicate that it is possible to use PCG to create content dynamically, adjusting it to the preferences of the user. This approach is known as experience-driven procedural content generation (EDPCG) (Raffe et al., 2015; Yannakakis & Togelius, 2011). In broad terms, EDPCG consists of an iterative process where the content is constantly modified based on the user's feedback.

The general functioning of EDPCG is the same as an evolutionary algorithm (EA), which is an optimization process inspired by natural evolution. In a natural environment, the organisms that are better adapted to their habitat tend to have more reproductive success, hence more likely to pass their genes to the next generation. Similarly, objects can be created programmatically in a virtual environment and tested to identify the most successful ones. The criteria to identify which objects are more successful is based on a previously defined goal. This goal is defined by the developer based on the purpose of the application. During each iteration, the objects that are more successful at reaching the goal are identified. In the following iterations, new sets of objects are created, and the characteristics of the most successful objects tend to remain, whereas the characteristics of the least successful tend to disappear. It is assumed that repeating this process several times allows to reach the optimal parameters required to achieve the goal. For example, if the goal is to create personalized visual representations of positive affective states, and the EA detects that the user tends to associate red, rounded objects with positive affective states, the game would produce objects that would tend to be more red and more angular. An introduction to EA can be found in Eiben and Smith (Eiben & Smith, 2015).

Additional research indicates that it is possible to create automatically visual compositions in VR using Deep Convolutional Neural Networks (DCNN) (Kitson et al., 2019). Overall, the process consists of merging features from two images to create a third image. This approach could be combined with EDPCG (Raffe et al., 2015; Yannakakis & Togelius, 2011) to create personalized affective visualizations. The process would involve at least three steps: (1) Create a set of VR content that all users will observe and used that content as a baseline. This initial set of content could be developed following the guidelines described in Table 1; (2) Capture user feedback about the visual stimuli to establish the affective state that each user associates to each piece of VR content; And (3) use DCNNs to merge features of the initial content onto new, personalized VR content.

## 3 Assessment of affective states

Users' feedback should be captured using methods that do not interrupt the VR experience, such as body movements (see Section 4.2) or electrophysiological signals, similar to Georgiou and Demiris (2017). Methods for assessing affective states can be grouped into three categories: self-report questionnaires, behavioral measures, and electrophysiological signals. Each method has advantages and disadvantages that will be discussed below.

### 3.1 Self-reports

Self-reports allow participants to evaluate their affective state by answering a series of questions. They can be used to verify the accuracy of the acquired information through other methods, such as behavioral and electrophysiological signals. Data collected through self-reports are often used as a ground-truth in the field of HCI.

In general, self-report measures are relatively easy to implement because they only require to display a series of questions on a paper sheet or a screen. Unlike behavioral and electrophysiological methods, self-reports are considered a direct measure because they allow asking participants directly about their mental states (Perkis et al., 2020). However, they are susceptible to be biased by rational processes. For example, participants who believe that it is expected from them to respond in a certain way might adjust their responses to fulfill that expectation, causing a phenomenon known as *experimenter bias* (Fisher, 1993). Some available tools for the assessment of affective responses are the Positive and Negative Affect Schedule (PANAS) (Watson et al., 1988), Self-Assessment Manikin (SAM) (Bradley & Lang, 1994), and Pick a Mood (PAM) (Desmet et al., 2016). The PANAS consists of 20 words related to negative and positive feelings (ten negatives and ten positives). Participants use those words to report their affective state. Each word can receive a rating from 1 to 5.

The SAM (Bradley & Lang, 1994) is an instrument that uses three scales: valence (pleasant / unpleasant), arousal (tension / relaxation) and dominance (inhibition / uninhibition). Each scale has five pictograms. Participants can select the blank spaces between each pictogram to indicate intermediate states. Therefore, answers to each scale can take values between 1 and 9 (see Figure 1). Given that this instrument is based on dimensions, it is compatible with dimensional models of affect. The SAM (Bradley & Lang, 1994) is one of the most established instruments for assessing affect (over 7.000 citations) and has been used for the development of batteries of stimuli with emotional content, such as the International Affective Pictures System (IAPS) (Lang et al., 2008) and the DEAP dataset (Koelstra et al., 2012).

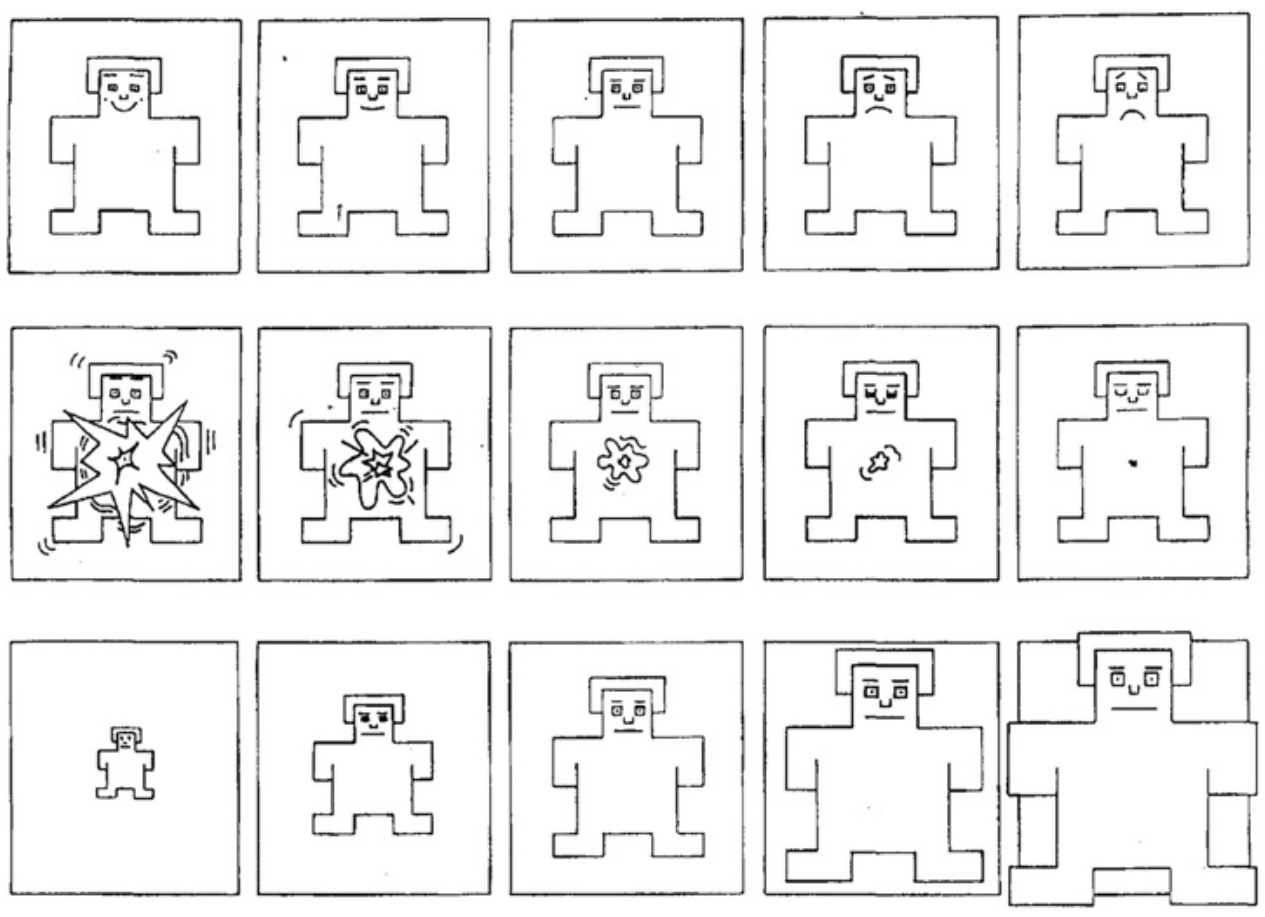

**Fig 1** From top to bottom: valence, arousal, and dominance scales of the Self-Assessment Manikin (SAM). Taken from (Bradley & Lang, 1994)

On the other hand, the PAM (Desmet et al., 2016) is based on discrete states. Therefore, it is compatible with discrete models of affect. This instrument also uses pictorial cues to assess participant's states. There are eight mood types plus a neutral one: excited, cheerful, relaxed, calm, bored, sad, irritated, and tense. There are three characters for each of these states: a man, a woman, and a robot (gender-neutral character). In comparison to the SAM (Bradley & Lang, 1994), PAM's characters (Desmet et al., 2016) are more similar to a real human being (see Figure 2), which might be an advantage because it could be easier for participants to feel identified with the characters of the PAM (Desmet et al., 2016).

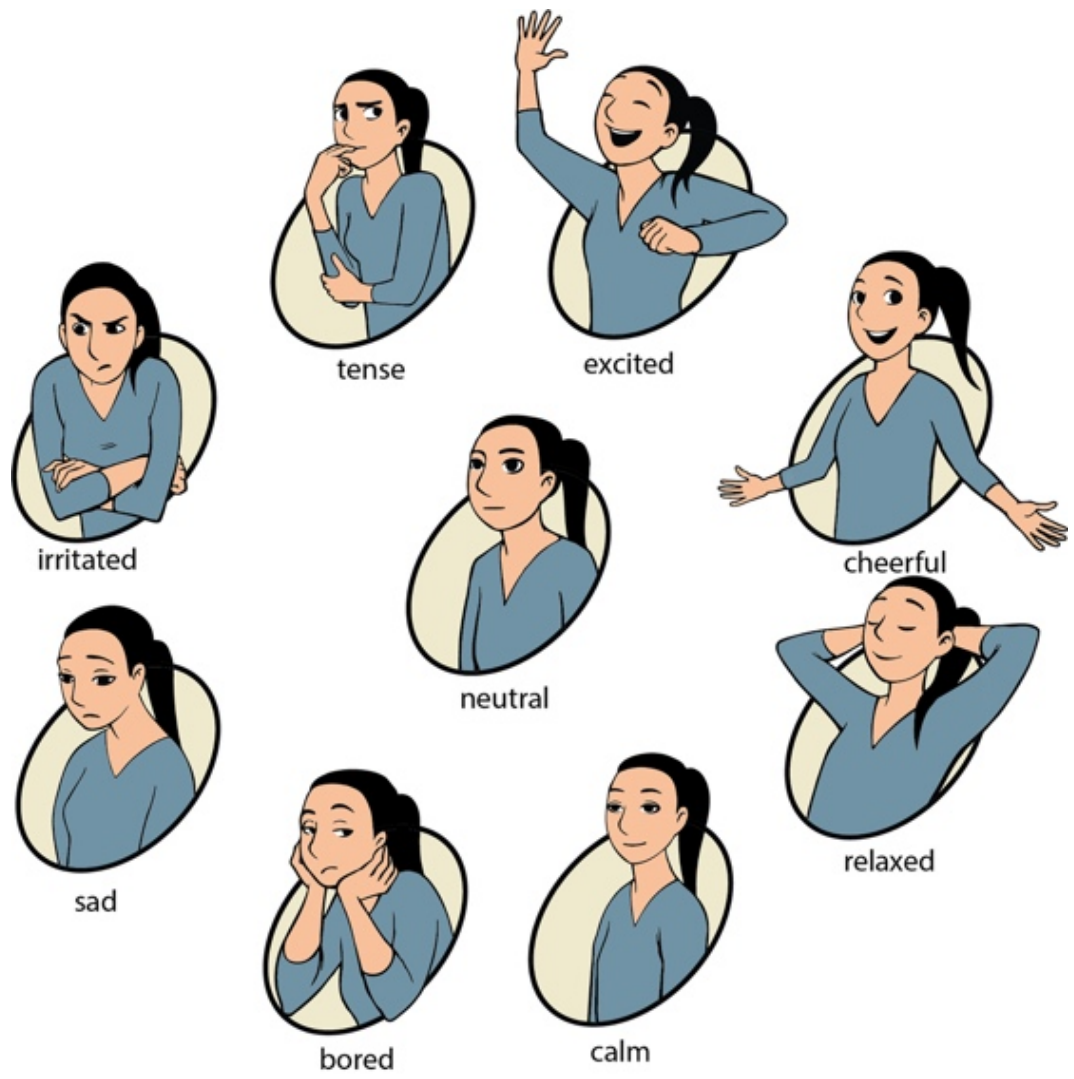

**Fig 2** Female character of Pick A Mood (PAM), taken from Desmet et al. (2016). Eight discrete states are represented, plus a neutral one.

The PAM has been used to understand how to design objects and experiences that could stimulate mood regulation (Desmet, 2015), analyze the effect of immersive virtual environments on gaming Quality of Experience (QoE) (Hupont et al., 2015), and analyze whether the effect of color on affective states varies across different VR rooms (Lipson-Smith et al., 2020). Using scales to analyze experiences in virtual environments might require to interrupt the VR experience. This limitation can potentially be counterbalanced by using subjective rating scales inside the virtual environment (Voigt-Antons et al., 2020).

## 3.2 **Behavioral measures**

Behavioral measures allow inferring affective states from observable conducts, such as body movements (Bull, 1978; Robitaille & McGuffin, 2019), voice patterns (Cordaro et al., 2016; Scherer & Oshinsky, 1977), and facial expressions (Ekman & Friesen, 1971). During an experiment conducted by Bull (1978), participants listened to a series of audio recordings with emotional content while their body movements were videotaped. Results suggested that sadness is associated with dropping the head while boredom is related to leaning the face in one hand. Building on that, recent research indicates that it is possible to infer arousal from body movements in virtual reality users (Kapur et al., 2005; Robitaille & McGuffin, 2019). In general, faster body movements are associated with higher arousal.

It is possible to automatically analyze users' affective states based on their voice patterns (Vogt et al., 2008). Usually, a set of features are defined and used to build a classification model. Some of the features used for automatic speech emotion recognition are pitch, loudness, and tempo (Polzehl et al., 2011; Vogt et al., 2008). This approach is coherent with evidence suggesting that changes in vocalization patterns have an effect on the affective evaluation of speech, e.g. (Banse & Scherer, 1996; Scherer & Oshinsky, 1977).

Eye-tracking has been an essential measure of various individual states or even personality traits (Hoppe et al., 2018). Greinacher & Voigt-Antons (2020) demonstrated recently how this measure could be easily obtained from modern smartphones using built-in system libraries (Greinacher & Voigt-Antons, 2020). The accuracy of this approach is comparable to other webcam or selfie-cam-based systems. However, as the authors pointed out, having eye-tracking systems easily accessible in millions of devices opens up opportunities for remote or in-the-field studies with a much higher ecological validity than studies relying on heavy equipment traditionally used in laboratory investigations.

As mentioned in section 3.1, facial expressions are associated with affective states (Ekman & Friesen, 1971). These expressions can be analyzed visually and described in terms of the Facial Action Coding System (FACS) (Ekman & Friesen, 1976). The FACS is an instrument that describes all the possible movements of the facial muscles. Each movement is defined as an Action Unit (AU). Facial expressions can be described as a combination of a subset of all the Action Units defined in the FACS (Ekman & Friesen, 1976). In a study conducted by Porcu et al. (2020), AUs were used for real-time analysis of the facial expressions of video streaming users. Additional studies suggest that human facial expressions can be collected using crowdsourcing techniques (D. McDuff et al., 2012), and its analysis can be optimized using statistical models that adapt automatically to the characteristics of the data (Feffer et al., 2018). However, facial recognition might be challenging to implement in VR because the Head-Mounted Display (HMD) covers the user's face. Therefore, facial electromyography (fEMG), a technique introduced in the following section, might be more suitable for capturing VR users' facial expressions.

## 3.3 Electrophysiology

Electrophysiological methods allow measuring changes in the electrical potentials of the body. Usually, facial electromyography (fEMG), electrocardiography (ECG), and electroencephalography (EEG) are used to record facial muscle, heart, and brain activity, respectively. This section focuses on methods to infer emotions in terms of the Circumplex Model of Affect (Russell, 1980) (see Section 2.2.). Therefore, the focus is on techniques that can be used to infer valence and arousal. There are many approaches for affect detection using electrophysiological signals that are not based on the Circumplex Model of Affect (Russell, 1980) and are not included in this manuscript.

Arousal can be inferred from features extracted from ECG signals. The beat-to-beat intervals of the ECG signal (often referred to as RR-Intervals, RRI) are extracted, detecting its peaks and calculating the time lapse between each peak. These RRIs are used to analyze the heart rate variability (HRV). Prominent examples of time-domain features used to analyze HRV are the root mean square of successive differences (RMSSD) and the standard deviation of NN intervals (SDNN). It has been found that higher HRV is associated with higher emotional arousal (Thayer et al., 2009). It is possible to extract features from the ECG signal in the frequency domain by calculating the LF/HF ratio. The low-frequency component (LF) (0.04 to 0.15 Hz) is associated with parasympathetic activity, while the high-frequency component (HF) (0.15 to 0.4 Hz) is associated with sympathetic activity (Malik et al., 1996). The activation of the parasympathetic system is associated with relaxation, and activation of the sympathetic system is associated with arousal. Therefore, more activity in the HF component indicates higher arousal (Pagani et al., 1984). Further research has shown that it is possible to infer arousal from EEG signals in VR users employing long short-term memory (LSTM) recurrent neural networks (RNN) (Hofmann et al., 2018).

A recent study compared the benefits of implementing HRV biofeedback in virtual reality with a traditional HRV biofeedback therapy (Blum et al., 2019), suggesting that the VR implementation produces more benefits for users in terms of relaxation self-efficacy, reduced mind wandering, and control of attentional resources. A similar approach was proposed in Blum et al. (2020), introducing a breathing biofeedback algorithm. This algorithm combines features extracted from electrocardiography activity with data inferred from diaphragm movements. The experiment was conducted using a chest band (Polar H10), which is a reliable, relatively inexpensive sensor. Results suggest that this approach can help to foster more regular and slower breathing in VR users.

Valence can be inferred from EMG and EEG signals. Previous evidence suggests that the Corrugator Supercilii muscle activity (located above the eyebrows) is associated with negative affective states. In contrast, the Zygomaticus Major muscle activity (located in the cheeks) is related to positive affective states (Dimberg, 1982). Changes in facial muscle activity can occur without conscious awareness of the participant (Dimberg et al., 2000; Dimberg & Thunberg, 2012). However, it might be challenging to implement EMG in a VR system because the pressure of the Head-Mounted Display (HDM) on the electrodes can create artifacts on the recorded signal.

Asymmetry in the cortical activity of the frontal cortex is also associated to valence . It has been found that positive and negative emotions are processed in the left and right frontal cortex, respectively (Huster et al., 2009; Ray & Cole, 1985; Antons, 2015). Additionally, it has been found that cortical activity decreases as the alpha power (frequencies between 8 and 13 Hz) increases (Pfurtscheller & Lopes da Silva, 1999). Therefore, increased processing of positive stimuli is associated with decreased alpha power in the left frontal cortex (higher activity in the left side of the brain). Similarly, increased processing of negative stimuli is associated with decreased alpha power in the right frontal cortex (higher activity in the right side of the brain) (Davidson, 1992; Huster et al., 2009; Pfurtscheller & Lopes da Silva, 1999).

These findings are coherent with results obtained by Reuderink et al. (2013) in a study where the brain activity of video game players was recorded using EEG. Participants were asked to report their affective state using the SAM (Bradley & Lang, 1994) after the game session ended. Results indicated a positive correlation between self-reported valence and alpha asymmetry. Likewise, Koelstra et al. (2012) analyzed the brain activity of 32 participants who watched forty musical videos and rated their emotional reactions to each video using the SAM (Bradley & Lang, 1994). A positive correlation was found between self-reported valence and alpha power in the right occipital region of the brain.

Eye-movements and eye-blinks cause artifacts in the EEG signals and are usually reflected in the activity of the frontal region of the brain. In non-stationary VR applications, it is particularly challenging to remove artifacts caused by muscle activity, head movements, or electrical activity from the VR headset (Klug & Gramann, 2020). It is possible to remove these artifacts using Independent Component Analysis (ICA). This technique allows to identify the components of an EEG signal that are not produced by brain activity (Makeig et al., 1997). The maximum number of independent components (ICs) that can be identified using ICA depends on the number of electrodes used. For example, a recording with 32 electrodes will allow to identify up to 32 ICs. Therefore, increasing the number of electrodes might help identify the artifacts in the signal with more precision. For a complete analysis about using ICA in non-stationary and stationary settings, see Klug and Gramman (2020).

An additional challenge is to process the EEG signals in real-time. ICA can be used in real-time (Pion-Tonachini et al., 2015), but it was not designed for that purpose. An alternative is Artifact Subspace Reconstruction (ASR) (S. Blum et al., 2019; Mullen et al., 2015), a technique designed for online artifact removal. ASR uses data recorded from the user as a baseline. Then, principal component analysis (PCP) is applied to identify the EEG channels that contain artifacts. The data of the corrupted channels are reconstructed using the baseline data as a reference. There is software available that can facilitate the implementation of ASR, such as BCILAB (Kothe & Makeig, 2013), OpenBiVE (Renard et al., 2010) and Neuropype (Intheon Labs, California).

### 3.4 Brain-Computer Interfaces

The implementation of electrophysiological signals in VR systems leads to the development of interfaces that allow interpreting users' brain activity as computer commands (Wolpaw et al., 2002). One of the basic assumptions underlying the development of Brain-Computer Interfaces (BCIs) is that mental processes originate in the brain. But there are BCIs that measure electrophysiological responses in other places of the body (e.g., Cassani et al., 2018), such as the heart and facial muscles, because cognitive processes that originate in the brain can produce changes in the activity of other body parts.

There are different techniques for measuring brain activity that can be used for the development of BCIs. For example, electrocorticography (ECoG), Positron Emission Tomography (PET), and functional Magnetic Resonance Imaging (fMRI), among others. However, electroencephalography (EEG) is the method most frequently used in BCIs because (1) it provides high temporal resolution (i.e., relatively large amount of data points recorded per second); (2) does not create health risks for the user because the electrodes can be easily placed and removed from the scalp; (3) can be portable, which is important for applications where the user is moving; and (4) is less expensive than most of the other methods (Zander & Kothe, 2011).

According to Zander & Kothe (2011), there are three types of BCIs: active, passive, and reactive. Active BCIs require the active participation of the user to generate an action. For example, patients who lack motor control can use mental commands to move a wheelchair (Voznenko et al., 2018). Passive BCIs do not require the conscious involvement of the user. They can be used, for example, to analyze the cognitive load of car drivers automatically (Almahasneh et al., 2014). Reactive BCIs use mental activity that occurs as a response to external stimuli. An example is a neurofeedback video game where threatening stimuli are presented, and players have to control their anxiety to obtain game score (Schoneveld et al., 2016). A VR application for affective visualization, would usually involve either a passive or a reactive BCI.

The typical workflow in a BCI involves at least four steps (Antons et al., 2014; Zander & Kothe, 2011):

1. **Preprocessing pipeline:** Filter out the signal's noise and keep only the components that reflect brain activity. This process involves (but is not limited to) filtering frequency bands and removing artifacts caused by eye-movements or muscle activity. An introduction to signal processing can be found in Unpingco (2014).
2. **Feature extraction:** Isolate the information related to the psychological construct of interest based on previous neuroscience studies (see Section 4.3).
3. **Classifier definition:** A classification model is created using prerecorded data. The classifier is tested offline, and an estimate of the accuracy of the classification is calculated. In general, classifiers are trained using data that has been previously labeled by humans. Machine Learning algorithms are used to identify patterns in the data that tend to be associated with each label.
4. **Classification application:** The classification is implemented in the BCI to perform online analysis of the brain activity. The outputs of the classification are used as computer commands.

## 4 Practical considerations

This section contains five practical considerations that might help during the development of a VR system for affective visualization.

1. **Which are the initial steps for designing a virtual environment?** First, define who will use the virtual environment (target group) and what the user will do inside that environment. This will help to have a more clear idea about the interaction events that will occur during the experience. Look for other interactive experiences, such as games and art installations, that can serve as inspiration. This will trigger ideas and will help to understand how to implement them. Then, define the graphical layout of your environment (color palette, typographies, and textures).
2. **Which software should I use for VR development?** Unity is probably one of the best options. There are alternatives, such as Vizard, a virtual reality software for research. However, to the extent of our knowledge, Unity is the only game engine compatible with open-source solutions, such as LSL and Excite-O-Meter. Therefore, it is relatively easy and inexpensive to develop virtual environments that rely on the user's physiological data using Unity.
3. **How to integrate Unity with electrophysiological equipment?** One possibility is to use LabStreamingLayer (LSL), a tool for collecting time-series data in experimental settings. Essentially, LSL allows to collect and synchronize the data and stream it into Unity. At the same time, it allows to send data from Unity (e.g., markers) to the signal processing software. Another option is to setup a UDP Broadcast to send information through your network.
4. **Is there a ready-to-use solution for integrating Unity with electrophysiological equipment?** Yes. Excite-O-Meter is a Unity plugin for visualizing cardiovascular activity, which is built on top of LSL. It can be used to visualize Heart Rate Variability (HRV) (see Section 4.3). By default, the Excite-O-Meter provides a time-series graph of the data. But you can customize it to build other types of visualizations.
5. **How to define the sampling rate for recording electrophysiological signals?** According the Nyquist-Shannon sampling theorem, the sampling rate should be twice the maximum frequency of interest. For example, if you are interested in frequencies of up to 128 Hz, you should use a sampling rate of at least 256 Hz. A sampling rate of 256 Hz means that you are collecting 256 data points per second.

   The usable information for each type of signal is located in a different frequency range. Therefore, the maximum frequency of interest for each signal is different. For example, the usable information in an ECG signal is up to 100 Hz. Therefore, the sampling frequency for ECG signals should be at least 200 Hz. However, previous studies indicate that ECG recordings at 200 Hz contain noise in the high-frequency components (Malik et al., 1996). This noise can be reduced by recording at a higher sampling rate. Therefore, it is considered a good practice to record ECG signals at a sampling rate between 256 Hz and 512 Hz, EMG signals at a sampling rate between 512 Hz and 1024, and EEG signals at a sampling rate between 256 Hz and 512 Hz.

## 5 Discussion

This manuscript aims to understand how to develop VR systems for affective visualization. These systems would involve the development of at least two components: a virtual environment and an affect detection technique. The development of both components requires the understanding of theories related to emotion and affect. Therefore, the manuscript analyses previous research related to (1) theories of emotion and affect, (2) audio-visual cues associated with affective states, and (3) methods for assessment of affective states.

Studies discussed in Section 3 suggest that specific visual and sound cues can represent users' emotions. However, most of these studies were conducted in experimental settings where the stimuli were carefully controlled. It is unclear whether the same psychological responses would occur if a combination of these cues were used simultaneously. For example, a particular combination of "happy" colors may result in an unbalanced visual composition that produces negative affective states. Or there might be motion patterns that are more prone to produce motion sickness in VR users, triggering negative states. Moreover, the novelty of a VR system in new users might bias the emotions they associate with the audio-visual stimuli.

Other studies mentioned in Section 3 suggest that leftwards linear motion tends to be associated with negative valence (Feng et al., 2014; Lockyer et al., 2011). This finding was obtained during experiments conducted in a western society, where time is represented as a progression to the right (Fuhrman & Boroditsky, 2010). Therefore, it is likely that western users associate leftward motion with negative affective states because that type of motion is culturally associated with *regressing* in time. However, in other cultures, such as the Hebrew culture, people represent time as a progression to the left (Fuhrman & Boroditsky, 2010). Therefore, it is possible that Hebrew users would associate leftward linear motion with positive valence. This hypothesis can be tested in future experiments.

Recent studies have demonstrated that affective states can be elicited by triggering psychogenic shivering (PS) (Haar et al., 2020; Schoeller, Haar, et al., 2019), using a device that controls the temperature in the upper back of the participants. Additional research indicates that the ability to be empathetic with others' emotions can be influenced by delivering electrical stimulation in the vagus nerve (Colzato et al., 2017), and by inducing affective states in the observer through videos (Pinilla et al., 2020). It remains an open question how to use those findings to develop Mixed Reality (MR) technologies for empathy enhancement, as proposed by Schoeller, Bertrand, et al. (2019).

Most of the existing techniques for inferring affective states from electrophysiological signals require the usage of previously annotated data to train a classifier, e.g. (Harischandra & Perera, 2012; Mavridou et al., 2017). But the amount of distinct affective states that can be detected using this approach is limited. Therefore, it might be convenient to formulate affect detection problems in terms of statistical regression. This approach would allow creating a model capable of describing affective states in terms of a continuum containing an infinite amount of distinct affective states. Previous studies suggest that it is possible to infer arousal from EEG signals (Hofmann et al., 2018) as a continuous variable. Future studies could investigate whether it is possible to use a similar approach to express valence in terms of a continuous variable.

Finally, it is possible to use a programmatic approach to create virtual reality content in real-time, using procedural content generation (PCG) (Bermudez i Badia et al., 2019; Raffe et al., 2015; Semertzidis et al., 2020; Yannakakis & Togelius, 2011). PCG allows to create content dynamically that adjusts to user feedback. Electrophysiological signals could be used to capture user feedback without interrupting the VR experience. This approach would allow to create personalized virtual environments for emotion visualization, similar to Kitson et al. (2019) or Bermudez i Badia (2019).

# 6 Declarations

## 6.1 Conflict of Interest

The authors declare that the research was conducted in the absence of any commercial or financial relationships that could be construed as a potential conflict of interest.

## 6.2 Author Contributions

All authors contributed to the study conception and analysis. Jaime Garcia and William Raffe contributed with analysis of data related to gaming and Virtual Reality. Jan-Niklas Voigt-Antons contributed to the analysis of data related to psychology and electrophysiology. Robert Philipp Greinacher contributed with the redaction of the manuscript and with data related to electrocardiography and eye-tracking. Sebastian Möller contributed to the analysis of data related to sound design and Machine Learning. Andres Pinilla performed the literature search, data analysis and wrote the first draft. All authors commented on previous versions of the manuscript.

### 6.3 Funding

We acknowledge the support of the German Research Foundation and the Open Access Publication Fund of TU Berlin.

### 6.4 Acknowledgments

This work was supported by the strategic partnership between the Technische Universität Berlin, Germany and the University of Technology Sydney, Australia.

We appreciate the generosity of Nick Busietta in giving us access to the Psydocs of LiminalVR (liminalvr.com). That documentation was crucial for writing Section 2 of this manuscript.

### 6.5 Data availability statement

Not applicable.

# 8 List of abbreviations

- Human Computer Interaction (HCI)
- Virtual Reality (VR)
- functional Magnetic Brain Imaging (fMRI)
- electroencephalography (EEG)
- Ortony, Clore & Collins (OCC) theory of emotions
- Evaluative Space Model (ESM)
- Positive and Negative Affect Schedule (PANAS)
- Self-Assessment Manikin (SAM)
- Pick a Mood (PAM)
- International Affective Pictures System (IAPS)
- Facial Action Coding System (FACS)
- Action Unit (AU)
- Head–Mounted Display (HMD)
- Electromyography (EMG)
- electrocardiography (ECG)
- electroencephalography (EEG)
- RR-Intervals (RRI)
- heart rate variability (HRV)
- root mean square of successive differences (RMSSD)
- standard deviation of NN intervals (SDNN).
- Low Frequency / High Frequency ratio (LF/HF ratio)
- long short-term memory recurrent neural networks (LSTMRNN)
- Independent Component Analysis (ICA)
- independent components (ICs)
- Artifact Subspace Reconstruction (ASR)
- principal component analysis (PCP)
- Brain–Computer Interfaces (BCIs)
- electrocorticography (ECoG),
- Positron Emission Tomography (PET)